\documentclass[twocolumn,showpacs,preprintnumbers]{revtex4}
\usepackage{graphicx}
\usepackage{amssymb}

\begin{document}
\title{Formation and evaporation of non-singular black holes}
\author{Sean A. Hayward}
\affiliation{Institute for Gravitational Physics and Geometry, The Pennsylvania
State University, University Park, PA 16802, U.S.A.}
\date{Revised 8th December 2005}

\begin{abstract}
Regular (non-singular) space-times are given which describe the formation of a 
(locally defined) black hole from an initial vacuum region, its quiescence as 
a static region, and its subsequent evaporation to a vacuum region. The static 
region is  Bardeen-like, supported by finite density and pressures, vanishing 
rapidly at large radius and behaving as a cosmological constant at small 
radius. The dynamic regions are Vaidya-like, with ingoing radiation of 
positive energy flux during collapse and negative energy flux during 
evaporation, the latter balanced by outgoing radiation of positive energy flux 
and a surface pressure at a pair creation surface. The black hole consists of 
a compact space-time region of trapped surfaces, with inner and outer 
boundaries which join circularly as a single smooth trapping horizon.
\end{abstract}
 \maketitle

{\em Introduction.} Black holes, predicted by Einstein gravity, appear to 
exist in the universe. The singularities which were also predicted to form 
inside them \cite{PH}, however, are generally regarded as indicating the 
breakdown of the theory, requiring modifications which presumably include 
quantum theory. A first step in this direction, quantum field theory on a 
stationary black-hole background, predicted Hawking radiation \cite{Haw1}. The 
ingoing radiation has negative energy flux which contradicts the assumptions 
of the singularity theorems and, in a semi-classical approximation, causes the 
black hole to shrink. In the usual picture \cite{Haw2}, the black hole shrinks 
until the central singularity is reached. However, if the singularity does not 
exist, such a picture cannot be correct \cite{AB,dis}.

\begin{figure}
\includegraphics[width=7cm]{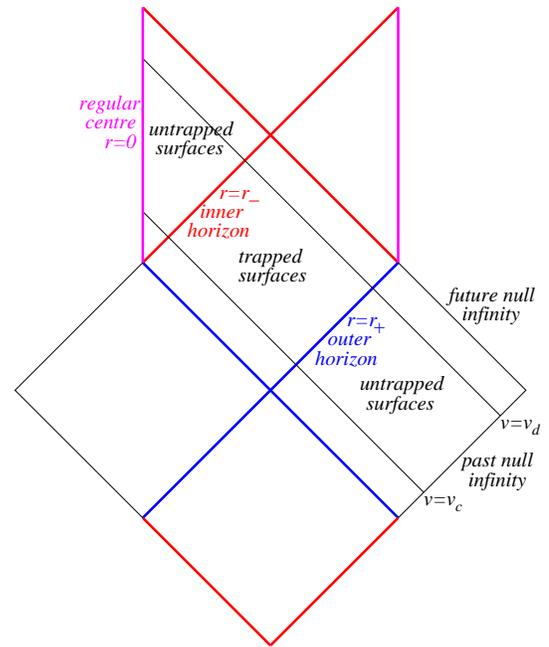}
\caption{Penrose diagram of a regular non-extreme black hole, to be identified 
vertically with isometric regions.} \label{bbh}
\end{figure}

Regular (i.e.\ non-singular) black holes have sometimes been considered, 
dating back at least to Bardeen \cite{Bar}. One can find metrics which are 
spherically symmetric, static, asymptotically flat, have regular centres, and 
for which the resulting Einstein tensor is physically reasonable, satisfying 
the weak energy condition and having components which are bounded and fall off 
appropriately at large distance. The simplest causal structure is similar to 
that of a Reissner-Nordstr\"om black hole, with the internal singularities 
replaced by regular centres (Fig.~\ref{bbh}). Such space-times have been 
dismissed as unphysical, due to the presence of a Cauchy horizon, but if such 
a black hole evaporates, the Cauchy horizon is no more real than the event 
horizon, as examples will show. 

Imagine removing astrophysically irrelevant regions to the past and future of 
two consecutive advanced times (Fig.~\ref{bbh}), then adjoining a past which 
describes gravitational collapse and a future which describes evaporation. The 
static region contains inner and outer horizons which no longer have global 
significance, but still have local significance as trapping horizons 
\cite{bhd}. The key issue is how the trapping horizons develop, which has been 
predicted on general principles \cite{dis}. In this Letter, concrete models 
are given for the collapse and evaporation phases, using Vaidya-like regions 
\cite{Vai} with ingoing or outgoing radiation.

\begin{figure}
\includegraphics[width=7cm]{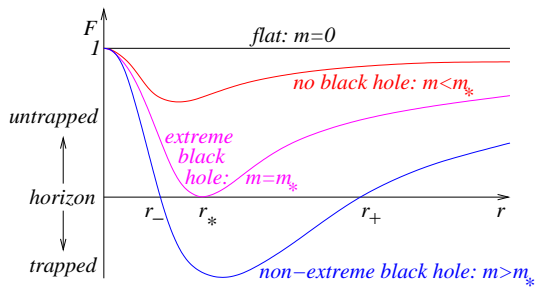}
\caption{The metric function $F=g^{rr}$, whose sign determines gravitational 
trapping, for fixed core radius $l$ and different total masses $m$.} \label{Fr}
\end{figure}

{\em Regular static black holes.} Consider static, spherically symmetric 
metrics of the form
\begin{equation}
ds^2=r^2dS^2+dr^2/F(r)-F(r)dt^2\label{g}
\end{equation}
where $t$ is the static time, $r$ the area radius and 
$dS^2=d\theta^2+d\phi^2\sin^2\theta$. A surface has area $4\pi r^2$, is 
trapped if $F(r)<0$ and untrapped if $F(r)>0$. Trapping horizons, in this case 
also Killing horizons, are located at the zeros $F(r)=0$, and there is a 
standard procedure to match regions across such horizons \cite{Wal}. For an 
asymptotically flat space-time with total mass $m$,
\begin{equation}
F(r)\sim1-2m/r\quad\hbox{as $r\to\infty$}.
\end{equation}
Similarly, flatness at the centre requires
\begin{equation} 
F(r)\sim1-r^2/l^2\quad\hbox{as $r\to0$}\label{F0}
\end{equation}
where $l$ is a convenient encoding of the central energy density $3/8\pi l^2$, 
assumed positive. A sketch of $F(r)$ indicating where it might dip below zero 
(Fig.~\ref{Fr}) shows that there will be a range of parameters for which there 
is no black hole, and that the simplest black-hole cases will generically have 
an inner and outer Killing horizon, the two cases separated by an extreme 
black hole with degenerate Killing horizon.

It can be shown that, for a metric $g$ of the form (\ref{g}), (\ref{F0}), the 
Einstein tensor has the cosmological-constant form 
\begin{equation} 
G\sim-\Lambda g\quad\hbox{as $r\to0$},\quad\Lambda=3/l^2.
\end{equation}
Thus there is an effective cosmological constant at small distances, with 
Hubble length $l$. Such behaviour has been proposed previously by Sakharov 
\cite{Sak} as the equation of state of matter at high density, and by Markov 
and others \cite{Mar} based on an upper limit on density or curvature, to be 
ultimately justified by a quantum theory of gravity. Since $l$ gives the 
approximate length scale below which such effects dominate, one might expect 
$l$ to be the Planck length or of the same order, though larger length scales 
are not excluded. 

{\em A minimal model.} For definiteness, take a particularly simple metric 
satisfying the above conditions:
\begin{equation}
F(r)=1-\frac{2mr^2}{r^3+2l^2m}\label{F}
\end{equation}
where $(l,m)$ are constants which will be assumed positive. This is similar to 
the Bardeen black hole, reduces to the Schwarzschild solution for $l=0$ and is 
flat for $m=0$. Poisson \& Israel \cite{PI} derived an equivalent form of 
$g^{rr}=F$ (without fixing $g_{tt}$) based on a simple relation between vacuum 
energy density and curvature.

\begin{figure}
\includegraphics[width=6cm]{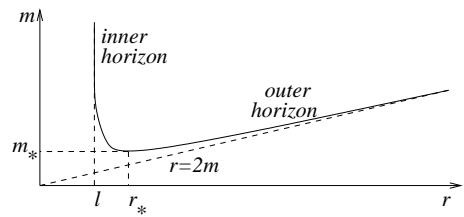}
\caption{Horizon mass-radius relation: a pair of horizons appears when mass 
$m$ exceeds critical mass $m_*$.} \label{mr}
\end{figure}

Elementary analysis of the zeros of $F(r)$ reveals a critical mass 
$m_*=(3\sqrt{3}/4)l$ and radius $r_*=\sqrt3l$ such that, for $r>0$,  $F(r)$ 
has no zeros if $m<m_*$, one double zero at $r=r_*$ if $m=m_*$, and two simple 
zeros at $r=r_\pm$ if $m>m_*$ (Fig.~\ref{Fr}). These cases therefore describe, 
respectively, a regular space-time with the same causal structure as flat 
space-time, a regular extreme black hole with degenerate Killing horizon, and 
a regular non-extreme black hole with both outer and inner Killing horizons, 
located at $r_+\approx 2m$ and $r_-\approx l$ for $m\gg m_*$ (Fig.~\ref{bbh}). 
The horizon radii $r_\pm$ determine the mass (Fig.~\ref{mr})
\begin{equation}
m(r_\pm)=\frac{\frac12r_\pm^3}{r_\pm^2-l^2}.\label{m}
\end{equation}
Note the existence of a mass gap: such black holes cannot form with mass 
$m<m_*$. Also, the inner horizon has radius $r_->l$ which is very close to $l$ 
for all but the smallest masses. In this sense, the black-hole core has a 
universal structure.

If the Einstein equation $G=8\pi T$ is used to interpret components of the 
energy tensor $T$, these metrics are supported by density $-T^t_t$, radial 
pressure $T^r_r$ and transverse pressure $T^\theta_\theta=T^\phi_\phi$ given by
\begin{eqnarray}
G^t_t=G^r_r&=&-\frac{12l^2m^2}{(r^3+2l^2m)^2}\label{Gn}\\
G^\theta_\theta=G^\phi_\phi&=&\frac{24(r^3-l^2m)l^2m^2}{(r^3+2l^2m)^3}.\label{Gt}
\end{eqnarray}
They fall off very rapidly, $O(r^{-6})$, at large distance. In terms of the 
energy $E$ defined by 
\begin{equation}
g^{rr}=1-2E/r\label{E}
\end{equation}
one finds the energy density $-T^t_t=(3l^2/2\pi)(E/r^3)^2$, proportional to 
the square of the curvature $E/r^3$. Poisson \& Israel \cite{PI} assumed such 
proportionality as a property of vacuum energy density; then the component 
$dE/dr=-4\pi r^2T^t_t$ of the Einstein equation implies $g^{rr}$ equivalent to 
(\ref{F}).

{\em Adding radiation.} Next rewrite the static space-times in terms of 
advanced time
\begin{equation}
v=t+\int\frac{dr}{F(r)}
\end{equation}
so that
\begin{equation}
ds^2=r^2dS^2+2dvdr-Fdv^2.
\end{equation}
Now allow the mass to depend on advanced time, $m(v)$, defining $F(r,v)$ by 
the same expression (\ref{F}). Then the density $-T^v_v$, radial pressure 
$T^r_r$ and transverse pressure $T^\theta_\theta$ have the same form 
(\ref{Gn})--(\ref{Gt}), but there is now an additional independent component, 
radially ingoing energy flux (or energy-momentum density) $T^r_v$ given by
\begin{equation}
G^r_v=\frac{2r^4m'}{(r^3+2l^2m)^2}
\end{equation}
where $m'=dm/dv$. This describes pure radiation, recovering the Vaidya 
solutions for $l=0$ and at large radius. In the Vaidya solutions, the ingoing 
radiation creates a central singularity, but in these models, the centre 
remains regular, with the same central energy density given by (\ref{F0}). It 
seems that the effective cosmological constant protects the core. 

\begin{figure}
\includegraphics[width=6cm]{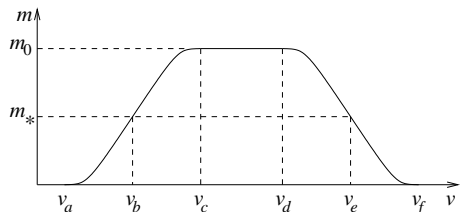}
\caption{A mass profile $m(v)$ in advanced time $v$.} \label{mv}
\end{figure}

The ingoing energy flux is positive if $m$ is increasing and negative if $m$ 
is decreasing. A key point is that trapping horizons still occur where the 
invariant $g^{rr}=F(r,v)$ vanishes \cite{bhd}. Then one can apply the previous 
analysis to locate the trapping horizons in $(v,r)$ coordinates parameterized 
by $m$, given by $m(r_\pm)$ in (\ref{m}) and a mass profile $m(v)$; 
qualitatively, by inspecting Figs.~\ref{mr} and \ref{mv}.

\begin{figure}
\includegraphics[width=85mm]{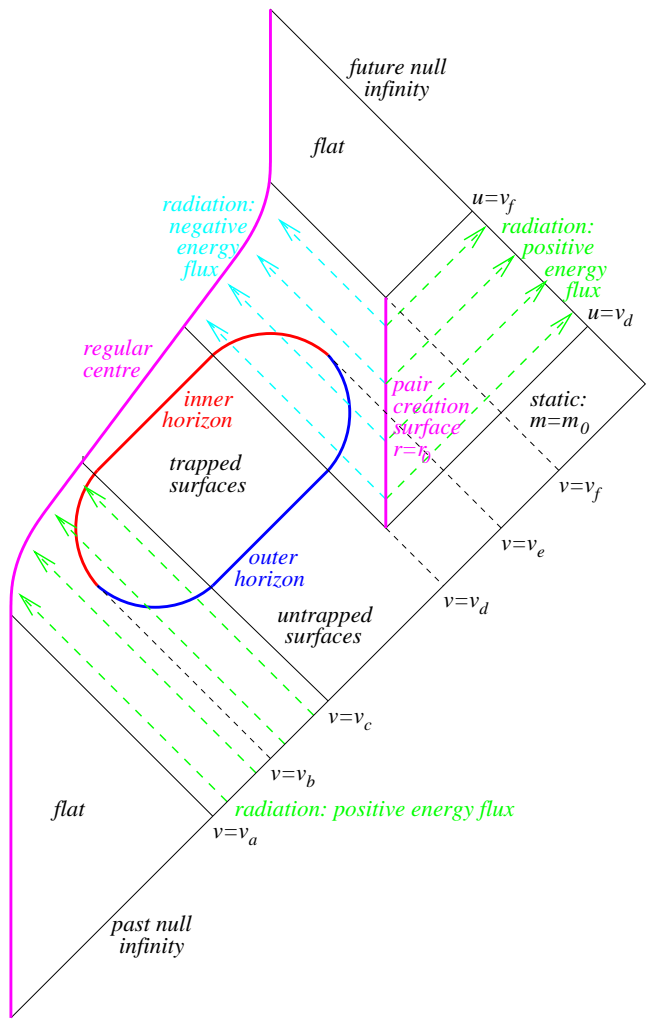}
\caption{Penrose diagram of formation and evaporation of a regular black hole 
in the given models.} \label{circus}
\end{figure}

{\em Ingoing radiation.} One can now model formation and evaporation of a 
static black-hole region. Introduce six consecutive advanced times 
$v_a<v_b<\ldots<v_f$ and consider smooth profiles of $m(v)$, meaning $m'(v)$ 
at least continuous, such that (Fig.~\ref{mv})
\begin{eqnarray}
\forall v\in(-\infty,v_a)&:&m(v)=0\label{va}\\
\forall v\in(v_a,v_c)&:&m'(v)>0\label{vb}\\
\forall v\in(v_c,v_d)&:&m(v)=m_0>m_*\label{vc}\\
\forall v\in(v_d,v_f)&:&m'(v)<0\label{vd}\\
\forall v\in(v_f,\infty)&:&m(v)=0.\label{ve}
\end{eqnarray}
Then
\begin{eqnarray}
\exists v_b\in(v_a,v_c)&:&m(v_b)=m_*\\
\exists v_e\in(v_d,v_f)&:&m(v_e)=m_*.
\end{eqnarray}
These transition times mark the appearance and disappearance of a pair of 
trapping horizons: for $v<v_b$ and $v>v_e$, there is no trapping horizon, 
while for $v_b<v<v_e$, there are outer and inner trapping horizons, in the 
sense of the author's local classification \cite{bhd}. These horizons join 
smoothly at the transitions and therefore unite as a single smooth trapping 
horizon enclosing a compact region of trapped surfaces (Fig.~\ref{circus}, for 
$r<r_0$).

{\em Outgoing radiation.} Thus far, only the ingoing Hawking radiation has 
been modelled, since outgoing radiation does not enter the equation of motion 
of the trapping horizon; in terms of retarded time $u$, $T_{vv}$ and $T_{uv}$ 
enter, but $T_{uu}$ does not \cite{bhd}. Outgoing Hawking radiation will now 
be modelled by adapting an idea of Hiscock \cite{His}: select a certain radius 
$r_0>2m_0$ outside the black hole, and adopt the above negative-energy 
radiation only inside that radius, balanced by outgoing positive-energy 
radiation outside that radius, with the same mass profile (Fig.~\ref{circus}). 
This is an idealized model of pair creation of ingoing particles with negative 
energy and outgoing particles with positive energy, locally conserving energy.

In more detail, consider an outgoing Vaidya-like region
\begin{equation}
ds^2=r^2dS^2-2dudr-Fdu^2
\end{equation}
with $F(r,u)$ as before (\ref{F}), with $m$ replaced by a mass function 
$n(u)$.  Fix the zero point of the retarded time $u$ so that $r=r_0$ 
corresponds to $u=v$. Now take the above model only for $v<v_d$ 
(\ref{va})--(\ref{vc}). For $v>v_d$, keep the profiles (\ref{vd})--(\ref{ve}) 
for $r<r_0$, but for $r>r_0$, take an outgoing Vaidya-like region with
\begin{eqnarray}
\forall u<v_d&:&n(u)=m_0\\
\forall u>v_d&:&n(u)=m(u).
\end{eqnarray}
Then there is a static region with total mass $m_0$ for $v>v_d$, $u<v_d$, and 
a flat region for $v>v_f$, $u>v_f$. Since the ingoing and outgoing radiation 
has no net energy but a net outward momentum, one might expect the pair 
creation surface $r=r_0$ to have a surface layer with no surface energy 
density but surface tension $\tau<0$. This is confirmed using the Israel  
formalism \cite{Isr}, yielding
\begin{equation}
-\frac{16\pi(g^{rr})^{3/2}}r\tau=[G^{rr}]=-\frac{4r^4m'}{(r^3+2l^2m)^2}
\end{equation}
at $r=r_0$, $v_d<v<v_f$.

The whole picture is given in Fig.~\ref{circus}. Action begins at $v=v_a$, a 
black hole begins to form at $v=v_b$, has collapsed completely at $v=v_c$ to a 
static state with mass $m_0$, begins to deflate at $v=v_d$ and eventually 
evaporates at $v=v_e$, leaving flat space finally after $v=v_f$, $u=v_f$. 
There is no singularity and no event horizon.

{\em Remarks.} A trapping horizon with both inner and outer sections typically 
develops in numerical simulations of binary black-hole coalescence, in 
analytical examples of gravitational collapse such as Oppenheimer-Snyder 
collapse and according to general arguments \cite{OS}. A key point here is 
that the inner horizon never reaches the centre, where a singularity would 
form \cite{dis}. This is compatible with the classical singularity theorems 
\cite{PH}, which make assumptions that are already not satisfied by a Bardeen 
black hole, such as the strong energy condition. The negative-energy nature of 
ingoing Hawking radiation shows that such theorems do not apply to a black 
hole that might someday begin to evaporate.

In contrast to the usual picture \cite{Haw2}, the endpoint $v=v_e$ of 
evaporation, defined locally by the disappearance of trapped surfaces, occurs 
when the outer and inner sections of the trapping horizon reunite. The 
subsequent timescale until the effective cessation of particle production at 
$v=v_f$ can be expected to be of the same order as $l$. Another logical 
possibility is that the inner and outer horizons approach each other 
asymptotically, forming the horizon of an extreme black hole with $m=m_*$, but 
such a delicately balanced situation would require justification.

The possibility of a circular trapping horizon has, in fact, been conjectured 
before \cite{FV}. Since there is no event horizon, long accepted as the 
defining property of a black hole, it seems necessary to stress that the 
static region looks just like a black hole over timescales that can be 
arbitrarily long. Thus it should be regarded as a black hole by any practical 
definition, as in the local, dynamical paradigm for black holes in terms of 
trapping horizons \cite{bhd}. The non-existence of an event horizon for a 
black hole which eventually evaporates seems to have been recently accepted by 
its most influential proponent \cite{GR17}.

Most discussions of black-hole evaporation mention a certain I-word, as a 
paradox, problem or puzzle. The above space-times, regular and with the causal 
structure of flat space-time, show that this word need not be mentioned. To 
paraphrase an old gravitational adage: what goes in, must come out.

\medskip 
Thanks to Abhay Ashtekar for discussions and James Bardeen for 
pointing out the surface layer. Research supported by NSF grants PHY-0090091, 
PHY-0354932 and the Eberly research funds of Penn State.

\end{document}